\documentclass[
aps,
prd,
reprint,
amsmath,
amssymb,
floatfix,
showkeys,showpacs,nofootinbib
]{revtex4-2}
\usepackage{amsmath,amssymb,bm}
\usepackage{graphicx}
\usepackage{hyperref}
\newcommand{\dd}{\mathrm{d}}
\newcommand{\eff}{\mathrm{eff}}

\newcommand{\vac}{\mathrm{vac}}
\newcommand{\nuc}{\mathrm{nuc}}
\newcommand{\false}{\mathrm{f}}
\newcommand{\true}{\mathrm{t}}
\newcommand{\calV}{\mathcal{V}}
\newcommand{\calJ}{\mathcal{J}}
\newcommand{\calU}{\mathcal{U}}

\begin{document}

\title{Thermal bubble nucleation with Gentile statistics}

\author{Orlando Luongo}
\email{orlando.luongo@unicam.it}
\affiliation{School of Science and Technology, University of Camerino, Via Madonna delle Carceri, Camerino, 62032, Italy.}
\affiliation{SUNY Polytechnic Institute, 13502 Utica, New York, USA.}
\affiliation{INAF - Osservatorio Astronomico di Brera, Milano, Italy.}
\affiliation{Istituto Nazionale di Fisica Nucleare (INFN), Sezione di Perugia, Perugia, 06123, Italy,}
\affiliation{Al-Farabi Kazakh National University, Al-Farabi av. 71, 050040 Almaty, Kazakhstan.}

\author{Hernando Quevedo}
\email{quevedo@nucleares.unam.mx}
\affiliation{Instituto de Ciencias Nucleares, Universidad Nacional Aut\'onoma de M\'exico, Mexico.}
\affiliation{Dipartimento di Fisica and Icra, Universit\`a di Roma ``La Sapienza'', Roma, Italy,}
\affiliation{Al-Farabi Kazakh National University, Al-Farabi av. 71, 050040 Almaty, Kazakhstan.}

\begin{abstract}
We investigate thermal bubble nucleation when a canonical scalar order parameter is coupled to an equilibrium particle bath obeying Gentile statistics. We derive an exact representation of the thermal free energy in terms of ordinary bosonic thermal integrals and show that the cubic mass contribution cancels for every finite occupation bound at high temperatures. Although this removes a conventional mechanism for generating metastability, it does not exclude first-order transitions. We consider a quartic scalar model with no prescribed barrier and construct its occupation-dependent phase diagram. When the vacuum contribution of the bath is omitted, the residual quartic logarithm can generate metastability and an exponentially small discontinuity in the order parameter. Including the conventional fermionic vacuum contribution restores a continuous transition near the symmetric spinodal in the fermionic limit. Formal nucleation and false vacuum conversion thresholds are reached before the spinodal, although the quasistatic approximation breaks down in the fermionic limit, while infrared scalar corrections indicate a loss of perturbative control for the first three occupation bounds. We also investigate a sextic model with an independently generated barrier, deriving analytical expressions for coexistence conditions and thermal properties. Increasing the occupation bound decreases the critical and nucleation temperatures while increasing the latent heat, interface tension and fractional supercooling. The resulting transitions remain weak and rapid. Finally, we discuss the contribution of a surviving relativistic Gentile population to the effective number of neutrino species.

\end{abstract}



\maketitle
\tableofcontents

\section{Introduction}\label{sec:intro}

Thermal phase transitions in the early universe depend on the 
response of the effective free energy to particle excitations 
and on whether a metastable phase can decay on cosmological 
time scales. For first order transitions, the equilibrium 
criterion of equal free energies differs from the dynamical 
conditions for bubble nucleation and false vacuum conversion. 
The common picture corresponds to a semiclassical description that involves critical 
bubbles, investigated through thermal fluctuation determinants and the competition 
between bubble growth and the universe expansion in a usual homogeneous and isotropic universe, see e.g. Refs.   
\cite{Langer1969,Affleck1981,Coleman1977,CallanColeman1977,
Linde1983,GuthWeinberg1981}. 

Clearly, this picture enters several 
extensions of the standard model of particle physics, involving additional scalar 
degrees of freedom and are relevant to the study of phase 
transitions in the very early universe \cite{AthronReview2024}.

The conventional treatment assumes to use either Bose-Einstein or Fermi-Dirac 
distributions, whose effects rely on different infrared properties, reflected 
in the thermal effective potential. In particular, a bosonic contribution can 
generate a nonanalytic cubic mass term and, so, a barrier 
between competing phases, whereas the analogous fermionic 
contribution has no cubic term. 

In an interacting theory, the 
bosonic contribution also depends on infrared screening and 
the treatment of thermal diagrams 
\cite{ColemanWeinberg1973,DolanJackiw1974,Dine1992,
ArnoldEspinosa1993}. 

In this respect, we wonder whether 
modifying the occupation statistics can remove 
the mechanism responsible for generating such a thermal 
barrier and whether a first order transition can still occur 
through other contributions to the effective potential.

Particularly, the so-called Gentile statistics impose a finite upper bound on the occupation 
of a single particle state\footnote{From Gentile picture, the Fermi-Dirac and Bose-Einstein 
distributions are recovered at the corresponding statistical limiting case \cite{Gentile1940,DaiXie2004,DaiXie2013}. For this reason, this alternative statistics are called intermediate, as they reproduce known results only as limiting case.} 

This 
prescription differs from fractional exclusion statistics, 
which modifies the counting of available many particle states 
\cite{Haldane1991,Wu1994}. For independent modes, the equilibrium 
partition function is well defined without modifying the 
canonical scalar field whose configurations describe the 
bubble walls. 

However, the occupation law alone does not determine 
the microscopic interactions of the additional particles. In this respect, the vacuum contribution or the dynamical processes 
responsible for maintaining thermal equilibrium are thus undetermined as well. 

Motivated by the above considerations, we here consider a 
phenomenological bath of such modes coupled to the scalar 
order parameter through a field dependent mass. In so doing, 
we can investigate how the occupation bound affects the 
thermal free energy while distinguishing the origin of the 
scalar barrier from the dynamics of bubble formation. We 
first derive the Gentile thermal contribution from its finite 
occupation partition function and express it exactly in 
terms of conventional bosonic thermal integrals. Expanding 
at high temperatures, we find that the cubic mass term 
cancels for every finite occupation bound. In particular, we show that the high temperature 
expansion is nonuniform in the occupation bound, so that 
the conventional cubic mass contribution is recovered at 
the bosonic limit. Further, we determine how the finite 
occupation bound modifies the thermal mass of the scalar 
order parameter. Even though this cancellation removes one source of metastability, 
other field dependent contributions to the thermal free energy 
may still generate a barrier. To this end, we examine a quartic 
scalar potential with a positive quartic coupling and no 
prescribed barrier. By solving the stationary point, 
coexistence and spinodal conditions using the full Gentile 
thermal integral, we determine whether metastable phases 
arise without inserting a cubic or sextic interaction term. 
Neglecting the vacuum contribution of the bath, we find that 
the remaining quartic logarithm can induce a discontinuity 
in the order parameter whose magnitude depends on the 
occupation bound and can become exponentially small. 
To characterize this behavior, we construct a phase 
diagram in which the magnitude of the discontinuity is 
investigated by varying the occupation bound and the 
couplings of the scalar sector. In addition, we examine the 
fermionic limiting case with its conventional vacuum contribution 
and show that the latter cancels the logarithmic 
field dependence of the thermal potential, recovering a 
continuous transition near the symmetric spinodal in our 
reference scenario. For the quartic potential, we also investigate nucleation 
and false vacuum conversion for our occupation 
bounds, since the coexistence conditions alone do not 
guarantee that the transition is completed. Assuming an adiabatically cooling cosmological 
background, we find that the formal nucleation, as well as percolation and conversion thresholds, are reached before the symmetric 
spinodal. However, the narrow metastability interval 
obtained for small occupation bounds motivates us to 
compare the characteristic properties, among which scales of bubble formation and  
scalar relaxation. In particular, 
we show that the quasistatic approximation breaks down at 
the fermionic limit. We also investigate the corrections 
to false vacuum conversion arising from the finite initial 
radius of the nucleated bubbles and, furthermore, we test the magnitude of infrared scalar corrections. Our estimates indicate that perturbative control is 
lost for the first three finite occupation bounds, whereas 
the local infrared expansion improves as the occupation 
bound increases. We then investigate how the occupation bound affects bubble 
nucleation when the scalar interactions themselves generate 
a barrier. For this purpose, we use an analytic sextic 
potential and, interestingly, within the approximation 
in which the Gentile contribution enters through the 
quadratic thermal mass correction, both the latent heat 
and the interface tension turn out to be independent of 
the occupation bound at fixed scalar potential parameters. Afterwards, we evaluate the complete thermal integrals in 
specified numerical bounds. In contrast to the aforementioned 
analytic approximation, the complete thermal integrals 
introduce an occupation dependence in the latent heat 
and interface tension. In particular, we find that 
increasing the occupation bound decreases the critical 
and nucleation temperatures, while increasing the latent 
heat, interface tension and fractional supercooling in 
our numerical benchmark. 

Last but not least, we also investigate how our numerical results depend 
on the assumptions entering the thermal nucleation rate and 
the velocity of the bubble walls. From a cosmological 
perspective, we discuss the strength and duration of the 
transitions obtained in our numerical benchmark and their 
possible implications for gravitational wave production. 
Finally, we evaluate how a surviving relativistic population 
obeying Gentile statistics could modify the effective 
number of neutrino species, discussing the role of its 
subsequent thermal history.

The paper is organized as follows. In Sec.~\ref{sec:background}, 
we derive the Gentile thermal free energy and investigate 
metastability and bubble nucleation for a quartic potential. 
We also examine the validity of the quasistatic 
approximation and estimate the infrared scalar corrections.
We also introduce the analytic sextic model. In 
Sec.~\ref{sec:numeric}, we numerically investigate bubble 
nucleation and false vacuum conversion for the sextic 
potential. In Sec.~\ref{sec:discussion}, we discuss the 
statistical limits, theoretical uncertainties and cosmological 
implications of our results. Finally, in 
Sec.~\ref{sec:conclusion}, we summarize our findings and 
outline the perspectives of our work.

\section{Background and Gentile statistics}\label{sec:background}

For a homogeneous scalar background, the equilibrium configurations follow from the extrema of the effective free energy density, $
 F(\phi,T)=-T\ln Z(\phi,T)$, where the partition function reads $ Z(\phi,T)=\operatorname{Tr} e^{-\hat H(\phi)/T}$, defined inside a physical volume $\calV$, providing an effective potential given by $
 V_{\eff}(\phi,T)=F(\phi,T)/\calV$. For conventional free bosonic and fermionic species with $E_p^2=p^2+m^2(\phi)$ and vanishing chemical potential\footnote{The conventional Euclidean construction uses periodic boundary conditions for bosons and antiperiodic boundary conditions for fermions. It produces Matsubara frequencies $2\pi nT$ and $(2n+1)\pi T$, respectively. No analogous frequency prescription will be assumed for the Gentile species.},
\begin{align}
f_{B;F}(m,T)&=\pm g_{B;F}T\!\int\!\frac{\dd^3p}{(2\pi)^3}
  \ln(1\mp e^{-E_p/T}),\label{fB}
\end{align}
With $y=m^2/T^2$, the standard bosonic thermal function is $
 J_B(y)=\int_0^\infty\!\dd x\,x^2
 \ln\!\left(1-e^{-\sqrt{x^2+y}}\right)$. The leading expansions for nonnegative masses are
\begin{subequations}
    \begin{align}
 f_B&=-\frac{g_B\pi^2T^4}{90}
  +\frac{g_Bm^2T^2}{24}
  -\frac{g_Bm^3T}{12\pi}+\cdots,\label{boson_exp}\\
 f_F&=-\frac{7g_F\pi^2T^4}{720}
  +\frac{g_Fm^2T^2}{48}+\cdots.\label{fermion_exp}
    \end{align}
\end{subequations}

The cubic contribution comes from the infrared sector of the bosonic thermal expression. Ring diagrams and thermal screening must be considered whenever this approximation is used for interacting conventional fields \cite{ArnoldEspinosa1993}.

To clarify the effect of the thermal mass, we consider the canonical quartic potential, plus the thermal scalar self correction,
$V_{\eff}^{\rm HT}\simeq \frac12\!\left(-\mu^2+\frac{\lambda T^2}{4}\right)\phi^2 +\frac{\lambda}{4}\phi^4$.
Within this approximation, the mean field critical scale is $T_0=2\mu/\sqrt\lambda$. Below it the nonzero minimum has $\phi_\pm^2=\mu^2/\lambda-T^2/4$. The minimum approaches the origin continuously as the temperature increases. 

For comparison, when a conventional bosonic sector induces an independent cubic interaction, the Landau potential takes the form \cite{ArnoldEspinosa1993}
\begin{equation}
 V_{\rm cub}=D(T^2-T_0^2)\phi^2
  -ET\phi^3+\frac\lambda4\phi^4.\label{cubic}
\end{equation}
Its coexistence relations  follow from $V'_{\rm cub}(\phi_c,T_c)=0$ and $V_{\rm cub}(\phi_c,T_c)=V_{\rm cub}(0,T_c)$, namely $
 \phi_c=\frac{2ET_c}{\lambda}$,
 $T_c^2=\frac{T_0^2}{1-E^2/(D\lambda)}$ and $
 V_{\rm cub}(\phi,T_c)=\frac\lambda4\phi^2(\phi-\phi_c)^2$.

{These relations require} $E^2<D\lambda$. The barrier is located at $\phi_c/2$ and has height $\lambda\phi_c^4/64$. {At coexistence,} its surface tension and latent heat are $\sigma_{\rm cub}=\frac{\phi_c^3}{6}\sqrt{\frac\lambda2}$ and $L_{\rm cub}=\frac{8E^2T_c^4}{\lambda^2}\left(D-\frac{E^2}{\lambda}\right)$.

\subsection{Gentile statistics and first order transitions}\label{sec:theory}\label{sec:q4scan}

The new sector, denoted by $X$, consists of independent modes whose occupancies are restricted to $0,1,\ldots,q$, for an integer $q\geq1$. We assume the relativistic dispersion $E_p=\sqrt{p^2+m_X^2(\phi)}$, an equilibrium temperature shared with the conventional plasma and zero chemical potential. These specifications define an ideal gas contribution to the effective free energy. They do not specify a fundamental interacting quantum field theory for $X$.

For $K=q+1$ and $x=e^{-E_p/T}$, a single mode contributes $ Z_{p,q}=\sum_{n=0}^{q}x^n
  =\frac{1-x^K}{1-x}$, having
\begin{align}
 n_q(E)&=x\frac{\partial\ln Z_{p,q}}{\partial x}
  =\frac{1}{e^{E/T}-1}
  -\frac{K}{e^{KE/T}-1}.\label{nq}
\end{align}
The equilibrium free energy is
\begin{equation}
 f_X^{(q)}(m,T)
 =\frac{g_XT}{2\pi^2}\int_0^\infty\dd p\,p^2
 \ln\!\left[\frac{1-e^{-\sqrt{p^2+m^2}/T}}
 {1-e^{-K\sqrt{p^2+m^2}/T}}\right].
 \label{fx_exact}
\end{equation}

Substituting $z=Kp/T$ in the second logarithm gives an exact identity,
\begin{subequations}
\begin{align}
 f_X^{(q)}(m,T)&=f_B^{(g_X)}(m,T)
 -Kf_B^{(g_X)}(m,T/K),\label{rescale}\\
 \calJ_q(y)&=J_B(y)-K^{-3}J_B(K^2y),\label{Jidentity}
\end{align}
\end{subequations}
where the superscript in $f_B^{(g_X)}$ specifies the degeneracy and $\calJ_q=2\pi^2f_X^{(q)}/(g_XT^4)$.

We {recover the statistical limits while preserving the canonical kinetic term of the scalar field}. At $q=1$, the mode partition function becomes $1+x$ and Eq.~\eqref{nq} is Fermi-Dirac. At fixed positive $E/T$, the limit $q\to\infty$ gives the Bose occupation. For finite $q$, the infrared expansion yields
$n_q(E)=\frac q2-\frac{K^2-1}{12}\frac ET+O_K\!\left(\frac{E^3}{T^3}\right)$.
The occupation therefore remains bounded as $E\to0$. {Consequently, the bosonic limit is nonuniform in the infrared regime.}

At vanishing mass, the standard integral $\int_0^\infty\dd x\,x/(e^x-1)=\pi^2/6$ gives $
 f_X^{(q)}(0,T)=-\frac{g_X\pi^2T^4}{90}
 \left(1-K^{-3}\right)$ and $\frac{\partial f_X^{(q)}}{\partial m^2}\big|_{m=0}
 =\frac{g_XT^2}{24}\left(1-K^{-1}\right)$.
For finite $K$ we may use the high temperature expansion of $J_B$ within $m/T\ll1$ and $Km/T\ll1$,
\begin{equation}
 J_B(y)=-\frac{\pi^4}{45}+\frac{\pi^2y}{12}
 -\frac{\pi y^{3/2}}6
 -\frac{y^2}{32}\ln\frac{y}{a_B}+O(y^3),
 \label{JB_series}
\end{equation}
where $a_B=16\pi^2e^{3/2-2\gamma_E}$.

The overall procedure leads to the interesting result for which the {cubic mass terms cancel in Eq.}~\eqref{Jidentity}, since $K^{-3}(K^2y)^{3/2}=y^{3/2}$, i.e.,
\begin{align}
 f_X^{(q)}(m,T)={}&-\frac{g_X\pi^2T^4}{90}(1-K^{-3})\nonumber\\
 &+\frac{g_Xm^2T^2}{24}(1-K^{-1})\nonumber\\
 &+\frac{g_Xm^4}{64\pi^2}\Bigl[(K-1)\ln\frac{m^2}{a_BT^2}
 \nonumber\\
 &\quad +2K\ln K\Bigr]+O_K\!\left(\frac{m^6}{T^2}\right).
 \label{HTgentile}
\end{align}

The cancellation of the cubic contribution raises a distinct question from
that investigated in the sextic benchmark: can a bath obeying Gentile
statistics support a metastable configuration when the tree potential has
no barrier? To address this point, we keep the canonical order parameter
and the field dependent mass already introduced above, set $m_{X0}=0$ and
consider the isolated $X$ thermal contribution\footnote{We deliberately omit
independent bosonic and fermionic thermal populations in this auxiliary
model, so as to identify the effect of the occupation law.}. The potential is
\begin{subequations}
    \begin{align}
 V_q^{(4)}(\phi,T)&=-\frac{\mu^2}{2}\phi^2
 +\frac{\lambda}{4}\phi^4+\Delta f_q(\phi,T),
 \label{q4V}\\
 \Delta f_q(\phi,T)&=f_X^{(q)}(h_X|\phi|,T)
 -f_X^{(q)}(0,T).
 \label{q4delta}
    \end{align}
\end{subequations}

\begin{figure*}[t]
\centering
\includegraphics[width=\textwidth]{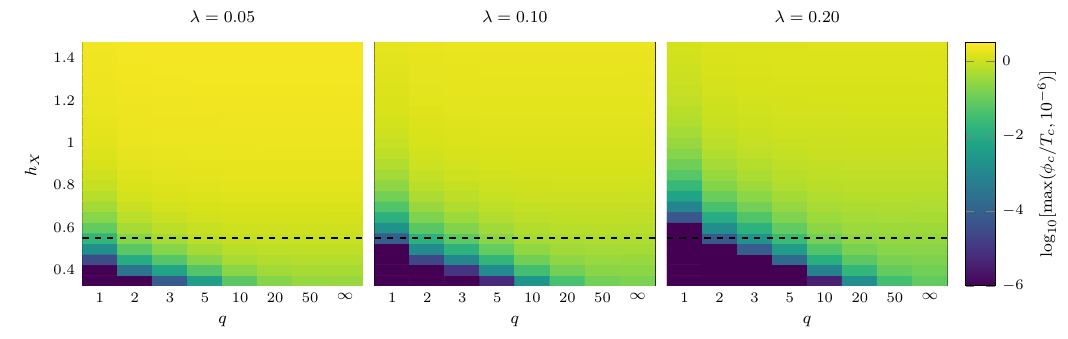}
\caption{Phase diagram of transition strength for the quartic thermal
only model of Eqs.~\eqref{q4V} and \eqref{q4delta}.
Colors encode $\log_{10}\!\left[\max(\phi_c/T_c,10^{-6})\right]$
for the three indicated positive quartic couplings. The occupation
bounds $q=1,2,3,5,10,20,50$ and $q\to\infty$ are treated as
separate statistical cases. The $h_X=0.55$ dashed line corresponds
to Table~\ref{tab:q4baseline}. A finite jump indicates coexistence
between the origin and a nonzero minimum. The color floor denotes an
exponentially weak jump, not a proof of a continuous transition.
The graph is obtained from the unexpanded equilibrium thermal integral,
with the stated omission of an $X$ vacuum contribution.}
\label{fig:q4phase}
\end{figure*}

We keep the renormalization prescription explicit: Eqs.~\eqref{q4V}
and \eqref{q4delta} define a \emph{thermal only} statistical model. No
vacuum determinant for $X$ has been postulated. Consequently,

\begin{itemize}
    \item[-] Defining $K=q+1$, $x=\phi/T$, $z=h_Xx$ and the exactly subtracted
thermal function
\begin{equation}
 \mathcal J_q(z)=J_B(z^2)-K^{-3}J_B(K^2z^2)
 +\frac{\pi^4}{45}(1-K^{-3}),
 \label{q4j}
\end{equation}
where $K^{-1}=0$ denotes the formal bosonic limit. Then
\begin{equation}
 \frac{V_q^{(4)}}{T^4}=-\frac{\mu^2}{2T^2}x^2
 +\frac{\lambda}{4}x^4
 +\frac{g_X}{2\pi^2}\mathcal J_q(h_Xx).
 \label{q4U}
\end{equation}
The symmetric extremum $x=0$ exists at every temperature. Its curvature
vanishes at
\begin{equation}
 T_0^2(q)=\frac{12\mu^2}{g_Xh_X^2(1-K^{-1})}.
 \label{q4t0}
\end{equation}
For every nonzero stationary point, differentiation of
Eq.~\eqref{q4U} gives $\frac{\mu^2}{T^2}=\mathcal A_q(z)$, with 
\begin{equation}
 \mathcal A_q(z)=\frac{\lambda z^2}{h_X^2}
 +\frac{g_Xh_X^2}{2\pi^2}
 \frac{\mathcal J'_q(z)}{z}.
 \label{q4stationary}
\end{equation}
Thus the full set of nonzero stationary points at any given $T$ is
obtained from the positive roots of
$\mathcal A_q(z)=\mu^2/T^2$. For increasing field, the first
root in the metastable interval is a local maximum and the second
is a broken minimum. This classification was checked by evaluating
the curvature and the free energies at the stationary points.

Coexistence requires both stationarity and equality of the free
energies. Eliminating $\mu^2/T_c^2$ between these conditions yields a
single dimensionless equation,
\begin{equation}
\frac{2\pi^2\lambda}{g_Xh_X^4}
 =\frac{4\mathcal J_q(z_c)-2z_c\mathcal J'_q(z_c)}{z_c^4},
 \label{q4coex}
\end{equation}
with $z_c>0$. Its solution determines
$\phi_c/T_c=z_c/h_X$ and $T_c=\mu\,[\mathcal A_q(z_c)]^{-1/2}$.
The lower spinodal coincides with Eq.~\eqref{q4t0}, while the
upper fold of the nonzero stationary branches satisfies $\mathcal A_q'(z_{\rm sp})=0$, with
\begin{equation}
 T_{\rm sp}=\frac{\mu}{\sqrt{\mathcal A_q(z_{\rm sp})}}.
 \label{q4fold}
\end{equation}
In the computed cases the ordering is
$T_0<T_c<T_{\rm sp}$, with the differences becoming exceedingly small
for weak transitions. These relations locate all stationary branches
without inserting a cubic interaction or assuming a transition in
advance.

\item[-] What replaces the cubic term?
At fixed finite $q$, the small mass expansion of Eq.~\eqref{q4j}
contains
\begin{align}
 \mathcal J_q(z)={}&\frac{\pi^2z^2}{12}(1-K^{-1})\nonumber\\
 &+\frac{z^4}{32}
 \left[q\ln\frac{z^2}{a_B}+2K\ln K\right]
 +O_q(z^6),\label{q4log}
\end{align}
with $a_B=16\pi^2\exp(3/2-2\gamma_E)$. The cubic mass term
is absent, while the quartic logarithm survives. In the particular
thermal only prescription, its coefficient in the potential is
\begin{equation}
 \Delta f_q\supset\frac{g_Xq h_X^4}{64\pi^2}
 \phi^4\ln\frac{h_X^2\phi^2}{a_BT^2}.
 \label{q4logv}
\end{equation}
This term can generate a very weak barrier, notwithstanding the
positive tree level quartic coefficient. In the domain $Kz_c\ll1$,
Eq.~\eqref{q4coex} gives the explicit estimate
\begin{equation}
 z_c\simeq\sqrt{a_B}\,
 \exp\!\left[-\frac{8\pi^2\lambda}{g_Xh_X^4q}
 -\frac{q+1}{q}\ln(q+1)-\frac12\right].
 \label{q4asym}
\end{equation}
The order parameter jump can therefore be exponentially suppressed,
without being strictly zero. The estimate applies for fixed finite
$q$, in the domain of the high temperature expansion; it does not
commute with the bosonic limit. The ordinary bosonic limit instead
contains the familiar negative cubic mass contribution.

\item[-] We choose $M=100\,{\rm GeV}$, $\mu^2/M^2=0.08$,
$g_X=8$, $h_X=0.55$ and $\lambda=0.10$ as a reference
quartic benchmark. Here $M$ sets units and is not assumed to be the
cutoff of an effective field theory. We evaluate the complete
integral with $q=1,2,3,5,10,20,50$ and the limiting bosonic
thermal function. At coexistence the three stationary configurations
are $\phi=0$, $\phi=\phi_{\rm bar}$ and
$\phi=\phi_c$, where the first and last are minima and the
intermediate one is a local maximum. The additional nonzero extrema
merge at $T_{\rm sp}$. At temperatures below $T_0$ the origin has
negative curvature, while above $T_{\rm sp}$ the nonzero stationary
branches disappear. Table~\ref{tab:q4baseline} reports the numerical
results. The small entries are quoted only at the precision supported
by the asymptotic treatment and quadrature.

To investigate parameter dependence, we scan
$h_X\in[0.35,1.45]$ and $\lambda=0.05,0.10,0.20$,
keeping $g_X$ and $\mu^2/M^2$ fixed. Figure~\ref{fig:q4phase}
shows the logarithm of the order parameter jump. Each column refers to
a \emph{discrete} occupation bound; neighboring columns are displayed
together for readability and do not imply a continuous statistical
interpolation. The horizontal dashed line identifies the reference
$h_X=0.55$. Values below $10^{-6}$ are shown at the same color
floor, but are not classified as continuous transitions.

The central finding is that suppressing the cubic mass term does not
remove metastability in this particular thermal only model. Finite
occupation bounds retain a quartic logarithm, which can yield a
first order transition with an exponentially small field jump and
barrier. The numerical scans do not reveal a sharp boundary between
continuous and discontinuous transitions for positive $h_X$ within
this prescription. In the decoupled limit $h_X=0$ the usual tree
quartic model recovers its continuous mean field transition.

This observation must be distinguished from a prediction of a
microscopic Gentile field theory. At $q=1$, for example, the ordinary
fermion vacuum determinant contains
\begin{equation}
 V_{{\rm vac},F}\supset
 -\frac{g_Xm_X^4}{64\pi^2}
 \left[\ln\frac{m_X^2}{\mu_R^2}-\frac32\right].
 \label{q4fermCW}
\end{equation}
Adding Eq.~\eqref{q4fermCW} to the thermal expression cancels
its $m_X^4\ln m_X^2$ dependence. The corresponding high temperature
quartic coefficient is
\begin{equation}
 \lambda_{\rm eff,F}(T)=\lambda+\frac{g_Xh_X^4}{16\pi^2}
 \left[\ln\frac{\mu_R^2}{a_BT^2}+4\ln2+\frac32\right].
 \label{q4matchedF}
\end{equation}
For $\mu_R=M$ and the reference parameters, this coefficient is
$\lambda_{\rm eff,F}(T_0)\simeq0.09581>0$. A direct scan of this
matched fermionic potential over $0\leq\phi/M\leq4$ in the
neighborhood of $T_0$ finds no competing broken minimum above $T_0$;
the broken solution approaches the origin continuously from below.
This limit check illustrates that the thermal only first order
behavior is not preserved under a physically motivated vacuum
completion. An analogous vacuum and infrared prescription is required
at the bosonic limit. No unique vacuum functional, gradient
correction or resummation follows from the Gentile occupation law at
intermediate $q$.
Therefore Fig.~\ref{fig:q4phase} establishes the phase structure of
the stated \emph{statistical thermal potential}; it does not prove
that a complete local quantum field theory has the same phase map.
A comparison between matched quantum actions will require additional
microscopic input.

The sextic construction considered next addresses the complementary
question: how does Gentile occupation modify the nucleation dynamics
when a barrier is already supplied by the scalar interactions?

\end{itemize}

\begin{table*}[t]
\caption{Quartic benchmark without a prescribed barrier, using the
exact Gentile thermal integral with no $X$ vacuum determinant.
The scales $T_0$, $T_c$ and $T_{\rm sp}$ and the dimensionless jump
$\phi_c/T_c$ are reported for $M=100\,{\rm GeV}$,
$\mu^2/M^2=0.08$, $g_X=8$, $h_X=0.55$ and $\lambda=0.10$.
The last column is the barrier height
$[V_q^{(4)}(\phi_{\rm bar},T_c)-V_q^{(4)}(0,T_c)]/T_c^4$.
The tiny separation of the three temperatures at $q=1$ is below the
displayed decimal precision.}
\label{tab:q4baseline}
\centering
\begin{tabular}{cccccc}
\hline\hline
$q$ & $T_0$ (GeV) & $T_c$ (GeV) & $T_{\rm sp}$ (GeV)
& $\phi_c/T_c$ & $\Delta V_{\rm bar}/T_c^4$\\
\hline
1 & 89.07235 & 89.07235 & 89.07235 & 8.52e-05 & $6.22e-21$ \\
2 & 77.13892 & 77.13920 & 77.13926 & 0.0144 & $1.02e-11$ \\
3 & 72.72727 & 72.73576 & 72.73756 & 0.0712 & $9.12e-09$ \\
5 & 68.99515 & 69.11334 & 69.13852 & 0.223 & $1.45e-06$ \\
10 & 66.05783 & 66.78809 & 66.94123 & 0.43 & $3.63e-05$ \\
20 & 64.53905 & 66.16402 & 66.47975 & 0.516 & $1.12e-04$ \\
50 & 63.61038 & 66.05198 & 66.44809 & 0.534 & $1.55e-04$ \\
$\infty$ & 62.98367 & 66.04366 & 66.44806 & 0.536 & $1.60e-04$ \\
\hline\hline
\end{tabular}
\end{table*}

\subsection{Thermal nucleation and conversion in the quartic model}
\label{sec:q4bounce}

The above analysis establishes a residual thermal barrier. We therefore investigate the same quartic
thermal potential of Eqs.~\eqref{q4V} and \eqref{q4delta},
without adding scalar interactions. Then, $O(3)$-symmetric critical configuration obeys
\begin{subequations}
    \begin{align}
 &\frac{\dd^2\phi}{\dd r^2}+\frac{2}{r}\frac{\dd\phi}{\dd r}
 =\frac{\partial V_q^{(4)}}{\partial\phi},
 \label{q4bounceode}\\
 &\frac{S_3^{(4)}}{T}=\frac{4\pi}{T}\int_0^\infty\!\dd r\,r^2
 \Bigl[\frac12\bigl(\phi'(r)\bigr)^2+V_q^{(4)}(\phi,T)-V_q^{(4)}(0,T)\Bigr].
 \label{q4bounceaction}
\end{align}
\end{subequations}
with $\phi'(0)=0$ and $\phi(\infty)=0$. We solve the radial
boundary problem for $T_0(q)<T<T_c(q)$ and determine the action
from the complete thermal potential. 

Afterwards, the surface tension
and latent heat are
\begin{subequations}
    \begin{align}
 \frac{\sigma_c}{T_c^3}&=\int_0^{\phi_c/T_c}\!\dd x\,
 \sqrt{\frac{2V_q^{(4)}(T_c x,T_c)}{T_c^4}},
 \label{q4sigma}\\
 \frac{L}{T_c^4}&=\frac{g_X}{2\pi^2}
 \left[4\mathcal J_q(z_c)-z_c\mathcal J'_q(z_c)\right].
 \label{q4L}
    \end{align}
\end{subequations}

Near coexistence, the thin wall limit gives
$S_3^{(4)}/T\simeq16\pi\sigma_c^3/
[3T_c L^2(1-T/T_c)^2]$. Nevertheless, for the background evolution we retain the spectator population
$g_{\rm sp}=100$, the prefactor $A(T)=T^4$, adiabatic cooling and
the wall speed $v_w=0.60$ employed in Sec.~\ref{sec:numeric}.
As the quartic auxiliary model contains no conventional $B,F$ bath,
its false-phase expansion is specified by
\begin{subequations}
    \begin{align}
 H_q^2(T)&=\frac{8\pi G}{3}\left[\rho_{\rm th}(T)
 +\frac{\mu^4}{4\lambda}\right],\\
 \rho_{\rm th}(T)&=\frac{\pi^2 T^4}{30}
 \left[g_{\rm sp}+g_X(1-K^{-3})\right].
 \label{q4H}
    \end{align}
\end{subequations}
We define $T_n$ by $N_H(T_n)=1$ using Eq.~\eqref{NH},
and use Eq.~\eqref{Iconversion} to obtain
$I(T_p)=0.34$ and $I(T_{95})=3$.

\begin{table*}[t]
\centering
\caption{{Quartic thermal-only benchmark with
$M=100\,\mathrm{GeV}$, $\mu^2/M^2=0.08$,
$g_X=8$, $h_X=0.55$, and $\lambda=0.10$.
Surface tensions and latent heats are normalized to $T_c$;
temperatures are in GeV. Nucleation and conversion assume
$g_{\rm sp}=100$, $A=T^4$, $T\propto a^{-1}$ and $v_w=0.60$.
The last row is the limiting bosonic thermal prescription.}}
\label{tab:q4dyn}
\begin{ruledtabular}
\begin{tabular}{cccccccc}
$q$ & $\sigma_c/T_c^3$ & $L/T_c^4$ & $T_n$ & $S_3(T_n)/T_n$
& $\beta/H$ & $T_p$ & $T_{95}$\\
$3$ & $6.33\times10^{-6}$ & $7.68\times10^{-4}$ & $72.73514$ & $130.11$ & $2.77\times10^{7}$ & $72.73497$ & $72.73496$\\
$5$ & $2.51\times10^{-4}$ & $8.35\times10^{-3}$ & $69.09936$ & $133.51$ & $1.13\times10^{6}$ & $69.09642$ & $69.09620$\\
$10$ & $2.41\times10^{-3}$ & $3.31\times10^{-2}$ & $66.68411$ & $135.69$ & $1.47\times10^{5}$ & $66.66645$ & $66.66500$\\
$20$ & $5.09\times10^{-3}$ & $4.86\times10^{-2}$ & $65.94864$ & $136.47$ & $7.02\times10^{4}$ & $65.91516$ & $65.91225$\\
$50$ & $6.25\times10^{-3}$ & $5.24\times10^{-2}$ & $65.78013$ & $136.72$ & $5.53\times10^{4}$ & $65.73890$ & $65.73524$\\
$\infty$ & $6.40\times10^{-3}$ & $5.27\times10^{-2}$ & $65.76269$ & $136.76$ & $5.33\times10^{4}$ & $65.72011$ & $65.71633$\\
\end{tabular}
\end{ruledtabular}
\end{table*}

Table~\ref{tab:q4dyn} shows that the thermal bounce reaches the
integrated nucleation criterion above the symmetric spinodal $T_0$
for all displayed cases. The operational percolation and $95\%$
conversion markers are  reached at $T>T_0$. In addition,
$\dot I>3H$ at $T_{95}$, so the physical false phase volume
decreases in our limit. 

We checked the bounce action
against its independent radial gradient (virial) representation. In particular, 
the relative discrepancy at onset is below $4\times10^{-6}$.
On the other side, the values of $\beta/H$ are large, indicating very short transitions
in Hubble units.

The much narrower coexistence windows at $q=1,2$ require a
numerically stabilized bounce shooting procedure. The corresponding
thermal solutions give $T_c-T_n\simeq2.34\times10^{-11}$ and
$9.53\times10^{-6}\,\mathrm{GeV}$ and $S_3(T_n)/T_n\simeq112.1$
and $125.6$, respectively. Although the formal conversion integrals reach the operational
thresholds before $T_0$ for both $q=1$ and $q=2$, the
quasistatic approximation breaks down for $q=1$, as shown
below. 

For $q=1,2$, we check the static bounce using Eq.~\eqref{q4logv}, with the
supercooling measured relative to coexistence in the same
approximation. For the most accurate solutions, the relative discrepancy with
the virial representation is below $10^{-10}$. The agreement with
the reported actions is within $0.1\%$. In particular, the matched fermionic vacuum contribution removes
the $q=1$ barrier as discussed above. Thus the dynamical calculation
gives formal nucleation and conversion criteria in the stated
\emph{thermal-only} quartic benchmarks, subject to the quasistatic
and infrared limitations discussed below. Despite the small latent
heat relative to the radiation density, reheating may become relevant. In particular, reheating and its effects on the coexistence interval are discussed in Ref.~\cite{Megevand2004}.

\subsection{Validity of the quasistatic nucleation regime}

The weak transitions obtained for small occupation bounds require an additional comparison between the spatial and temporal scales associated with bubble formation. We define the characteristic radius of the critical configuration as

\begin{equation}
R_b(T)=\underset{r}{\operatorname{argmax}}\,
|\partial_r\varphi(r,T)|.
\end{equation}

The time available for nucleation before the symmetric phase loses its local stability is

\begin{equation}
\tau_{\rm meta}=
\int_{T_0}^{T_c}\frac{dT}{TH_q(T)}.
\end{equation}

For sufficiently weak transitions, the logarithmic potential following from Eq.~\eqref{q4logv} can be written at coexistence as

\begin{equation}
V_q^{(4)}(\varphi,T_c)=B_q
\left[
\varphi_c^2\varphi^2+
\varphi^4
\left(
\ln\frac{\varphi^2}{\varphi_c^2}-1
\right)
\right],
\label{q4coexlog}
\end{equation}

where $B_q=g_Xqh_X^4/(64\pi^2)$. Accordingly, the coexistence surface tension becomes

\begin{equation}
\begin{aligned}
\sigma_c&=\sqrt{2B_q}\varphi_c^3
\int_0^1dy\sqrt{y^2+y^4(\ln y^2-1)}\\
&\simeq0.296914\sqrt{B_q}\varphi_c^3.
\end{aligned}
\end{equation}

For $q=1$, we obtain $\sigma_c\simeq4.42\times10^{-9}\,\mathrm{GeV}^3$ and $T_c-T_0\simeq7.43\times10^{-9}\,\mathrm{GeV}$. The thin wall estimate gives $R_c(T_n)\simeq7.30\times10^5\,\mathrm{GeV}^{-1}$, in agreement with the numerical radius $R_b(T_n)\simeq7.33\times10^5\,\mathrm{GeV}^{-1}$ obtained from the logarithmic bounce. The corresponding action is $S_3(T_n)/T_n\simeq112.19$.

However, the metastability interval lasts only $\tau_{\rm meta}\simeq7.48\times10^3\,\mathrm{GeV}^{-1}$, while $\beta^{-1}\simeq0.106\,\mathrm{GeV}^{-1}$. Consequently, the separation between the bubble formation scale and the time scale over which the nucleation rate changes is lost.

A complementary estimate follows from the curvature of the false phase,

\begin{equation}
m_f^2(T)=
\left.
\frac{\partial^2V_q^{(4)}}{\partial\varphi^2}
\right|_{\varphi=0}
=C_q(T^2-T_0^2),
\end{equation}

where $C_q=g_Xh_X^2q/[12(q+1)]$.

Within a local dissipative description, the homogeneous scalar fluctuations satisfy

\begin{equation}
\delta\ddot{\varphi}+D_q\delta\dot{\varphi}
+m_f^2\delta\varphi=0,
\end{equation}

with $D_q=3H+\gamma_q$ and $\gamma_q$ a phenomenological damping coefficient. The minimum relaxation time within this description is obtained at critical damping, $D_q=2m_f$, yielding $\tau_{\rm rel}^{\min}=m_f^{-1}$.

The condition $\beta/m_f\ll1$ therefore provides a necessary test for quasistatic relaxation within this prescription. For $q=1$, we obtain $\beta/m_f\simeq2.60\times10^4$, whereas for $q=2$ the corresponding ratio decreases to $2.16\times10^{-4}$. For the remaining occupation bounds considered in Table~\ref{tab:q4dyn}, this ratio is smaller than $5\times10^{-7}$.

The relaxation condition can be satisfied for $\beta<m_f$ provided that the damping coefficient belongs to the interval

\begin{equation}
2\beta<D_q<\beta+\frac{m_f^2}{\beta}.
\end{equation}

The occupation law does not determine $\gamma_q$. Thus, the above relation constrains the dynamical realization of the equilibrium model rather than specifying its relaxation rate.

We further examine the approximation of negligible initial bubble radius adopted in the conversion calculation. Including the physical radius at nucleation modifies Eq.~\eqref{Iconversion} according to

\begin{equation}
\begin{aligned}
I_R(T)&=\frac{4\pi}{3}\int_T^{T_c}
\frac{dT'\,\Gamma(T')}{H(T')T'^4}\\
&\quad\times\left[
T'R_b(T')+
v_w\int_T^{T'}\frac{dT''}{H(T'')}
\right]^3.
\end{aligned}
\end{equation}

For a rapidly varying nucleation rate, $\Gamma(t')\simeq\Gamma(t)e^{-\beta(t-t')}$, neglecting the variation of $H$ and $R_b$ over the characteristic nucleation interval gives

\begin{equation}
\frac{I_R}{I_0}
=1+\epsilon_R+\frac{\epsilon_R^2}{2}
+\frac{\epsilon_R^3}{6},
\qquad
\epsilon_R=\frac{\beta R_b}{v_w},
\end{equation}

where $I_0$ denotes the conversion exponent obtained by neglecting the initial radius.

For $q=2$, the numerical bounce yields $R_b(T_n)\simeq333\,\mathrm{GeV}^{-1}$ and $\beta/H\simeq1.91\times10^9$, giving $\epsilon_R\simeq8.93\times10^{-3}$. Including the initial radius in the conversion integral shifts the operational percolation and conversion temperatures by approximately $2.67\times10^{-10}$ and $2.63\times10^{-10}\,\mathrm{GeV}$, respectively, within the logarithmic potential and prescribed cooling history.

For $q\geq3$, we obtain $\epsilon_R<10^{-5}$ at nucleation. Consequently, the initial bubble radius produces negligible corrections at the numerical precision of Table~\ref{tab:q4dyn}.

The $q=1$ thermal-only benchmark instead violates the spatial and temporal hierarchies required by the quasistatic prescription. Its static bounce remains mathematically defined but the corresponding nucleation and conversion temperatures cannot be interpreted as controlled dynamical predictions. Establishing the evolution of this regime requires a time dependent description of the scalar fluctuations and their coupling to the thermal environment.

For reference, the characteristic radius decreases from
$R_b(T_n)\simeq28.5\,\mathrm{GeV}^{-1}$ at $q=3$ to
$0.981\,\mathrm{GeV}^{-1}$ at the bosonic thermal limit, while
$\beta^{-1}$ increases from $4.83\times10^6$ to
$3.07\times10^9\,\mathrm{GeV}^{-1}$. Over the same range,
$\beta/m_f$ decreases from $4.98\times10^{-7}$ to
$3.84\times10^{-11}$ and $\epsilon_R$ from $9.85\times10^{-6}$
to $5.33\times10^{-10}$.

\noindent{Infrared scalar fluctuations.}
The quasistatic conditions do not establish the validity of the
scalar loop expansion. At the broken minimum $(\phi_c,T_c)$, we define $
m_b^2=\partial_\phi^2 V_q^{(4)}$, $
g_3=\partial_\phi^3 V_q^{(4)}$ and $g_4=\partial_\phi^4 V_q^{(4)}$, where $m_b>0$. The separation of soft and hard thermal modes is discussed in Ref.~\cite{Kajantie1996}. The leading infrared contribution of the canonical
scalar zero mode, treating the specified potential as the input
for the soft scalar sector, is
$\Delta V_{\rm IR}=-T[\partial_\phi^2 V_q^{(4)}]^{3/2}/(12\pi)$
in the positive curvature region. Its second derivative at the
uncorrected minimum gives the local curvature correction
\begin{equation}
\Delta V_{\rm IR}''
=-\frac{T_c g_4 m_b}{8\pi}
-\frac{T_c g_3^2}{16\pi m_b}.
\end{equation}
We therefore define
\begin{subequations}
\begin{align}
&\epsilon_3=\frac{T_c g_3^2}{16\pi m_b^3},\\
&\epsilon_4=\frac{T_c |g_4|}{8\pi m_b}.
\label{q4IRparameters}
\end{align}
\end{subequations}
These quantities estimate the size of infrared corrections relative
to the input curvature without shifting the minimum or including
the hard thermal modes. Perturbative control requires both to be
much smaller than unity.

Differentiating Eq.~\eqref{q4coexlog} yields
$m_b^2=4B_q\phi_c^2$, $g_3=28B_q\phi_c$ and $g_4=76B_q$.
In the logarithmic regime we thus obtain
\begin{subequations}
    \begin{align}
&\epsilon_3=\frac{49}{8\pi}\frac{\sqrt{B_q}}{\phi_c/T_c},\\
&\epsilon_4=\frac{19}{4\pi}\frac{\sqrt{B_q}}{\phi_c/T_c}.
\label{q4IRlog}
    \end{align}
\end{subequations}
Consequently, the exponential suppression of the field
discontinuity in Eq.~\eqref{q4asym} can enhance the infrared
corrections exponentially. To evaluate the local expansion
parameters, for $q=1$ we use Eq.~\eqref{q4IRlog},
whereas for $q\geq2$ we differentiate the complete thermal
function using the convergent representation
$J_B(z^2)=-z^2\sum_{n=1}^\infty K_2(nz)/n^2$ at positive $z$,
where $K_2$ denotes the modified Bessel function.

For $q=1,2,3$, the pairs $(\epsilon_3,\epsilon_4)$ are
approximately $(779,604)$, $(6.51,5.05)$ and $(1.61,1.25)$,
respectively. At $q=5$ they decrease to $(0.655,0.505)$,
whereas for $q=10,20,50$ we obtain $(0.423,0.309)$,
$(0.356,0.236)$ and $(0.326,0.212)$. The bosonic thermal
limit gives $(0.323,0.211)$. The ratio $m_b/T_c$ increases
from $5.80\times10^{-6}$ at $q=1$ to $0.133$ at the bosonic
thermal limit.

The expansion is not controlled for $q=1,2,3$ and improves as the
occupation bound increases, although even at the bosonic thermal
limit the local parameters are not much smaller than unity.
These estimates do not establish the stability of the barrier or
perturbative control along the entire bounce profile. The weak
quartic transitions must therefore be interpreted within the
specified scalar potential. 

A nonperturbative treatment of the
scalar sector and its matching to the thermal bath will be explored in
future works, comparing our future findings with previous literature, see e.g. Refs.  \cite{Gould2021,GouldHirvonen2021,MooreRummukainen2001,MooreTranberg2001}. 

\subsection{Thermal mass correction and an analytic sextic model}

Alternatively to the previous description, we can here assume a barrier that is phenomenologically given by a sixth order potential, following the standard prescription \cite{GrojeanServantWells2005}, $m_X^2(\phi)=m_{X0}^2+h_X^2\phi^2$

For $m_{X0}=0$ and $Kh_X|\phi|/T\ll1$, Eq.~\eqref{HTgentile} implies
$\Delta V_{T,X}^{(q)}\supset \frac{g_Xh_X^2T^2}{24}\frac{q}{q+1}\phi^2$ and $\Delta m_{\phi,X}^2(T)=\frac{g_Xh_X^2T^2}{12}\frac{q}{q+1}$.

{We now consider a first order transition whose scalar barrier is independent of the occupation law.} Consider the Landau effective potential
\begin{align}
 V_{\rm an}(\phi,T)=\frac12 a_q(T)\phi^2
 -\frac b4\phi^4+\frac c6\phi^6,\label{sextic},
\end{align}
with $a_q(T)=-\mu^2+C_qT^2$ and $
 C_q=C_0+\frac{g_Xh_X^2}{12}\frac q{q+1}$.

{The potential preserves the} $\mathbb Z_2$ symmetry {of a real scalar}, and the analysis may be restricted to $\phi\ge0$. We regard the coefficients as phenomenological parameters, including the dimensions implicit in $c$. Stationary configurations obey $
 \phi\left(a_q-b\phi^2+c\phi^4\right)=0$ and
 $u_\pm=\phi_\pm^2=
 \frac{b\pm\sqrt{b^2-4ca_q}}{2c}$.

 When $0<a_q<b^2/(4c)$, $u_-$ is the barrier and $u_+$ is the broken minimum. Simultaneous stationarity and equality of free energies at coexistence yield $
 u_c\equiv\phi_c^2=\frac{3b}{4c}$,
 $a_c=\frac{3b^2}{16c},\qquad
 T_c^2=\frac{\mu^2+a_c}{C_q}$, $
 V_{\rm an}(\phi,T_c)=
 \frac c6\phi^2(\phi^2-u_c)^2$.

Thus $\phi_{\rm bar}^2=u_c/3$, the barrier height is $b^3/(96c^2)$ and the two spinodal scales are
$T_0^2=\frac{\mu^2}{C_q}$ and $T_{\rm sp}^2=\frac{\mu^2+b^2/(4c)}{C_q}$.
The ordering is $T_0<T_c<T_{\rm sp}$. {Within this approximation,} the occupation dependence enters through $C_q$, while the barrier at coexistence retains the same algebraic shape.

For a phase minimum $\phi_i(T)$, the entropy density is $
 s_i=-\frac{\dd f_i}{\dd T}$, and energy densities are $
 \rho_i=f_i-T\frac{\dd f_i}{\dd T}$, with $f_i=V_{\rm an}(\phi_i,T)$. Since $V_{\rm an}'(\phi_i,T)=0$, the total derivative reduces to the partial thermal derivative.

Hence, invoking  $\Delta f=f_{\false}-f_{\true}$ and $\Delta\rho=\rho_{\false}-\rho_{\true}$ and recalling that $\Delta\rho=\Delta f-T\frac{\dd\Delta f}{\dd T}$ and $L=\Delta\rho(T_c)=T_c
 \left.\frac{\partial V_{\rm an}}{\partial T}\right|_{\phi_c,T_c}=C_qT_c^2u_c$, we obtain
$L=\left(\mu^2+\frac{3b^2}{16c}\right)\frac{3b}{4c}$.
{At this order,} $L$ is independent of $q$ for fixed $\mu,b,c$. This property is not expected to hold exactly once the full nonquadratic thermal functions are introduced.

The {corresponding} wall profile, increasing from the symmetric to the broken phase, is\footnote{The planar wall at coexistence obeys $\phi''=\partial V_{\rm an}/\partial\phi$. Its first integral and tension give
$\frac12\left(\frac{\dd\phi}{\dd z}\right)^2=V_{\rm an}(\phi,T_c)$
with $\sigma_c=\int_0^{\sqrt{u_c}}\!\dd\phi\,
 \sqrt{2V_{\rm an}(\phi,T_c)}$ that gives $
 \sigma_c=\frac{9b^2}{64\sqrt3\,c^{3/2}}$.}
\begin{equation}
 \phi^2(z)=\frac{u_c}
 {1+\exp[-2u_c\sqrt{c/3}\,(z-z_0)]}.
 \label{wall}
\end{equation}
The derivation is exact for the analytic potential at coexistence. Away from this point, the profile and wall tension generally require a separate boundary value calculation.

In the thin wall limit, the bubble energy for radius $R$ is
$F(R)=4\pi R^2\sigma-\frac{4\pi}3R^3\Delta f$,
where $R_c=\frac{2\sigma}{\Delta f}$ and $ F_c=\frac{16\pi\sigma^3}{3\Delta f^2}$.

The critical configuration is unstable under radial changes because $F''(R_c)=-8\pi\sigma<0$. Close to $T_c$, $\Delta f=L(1-T/T_c)+\ldots$, so the corresponding thin wall exponent diverges as the inverse square of the supercooling. The cosmological benchmark below is evaluated with a numerical bounce, without identifying $F_c$ with the exact action away from thin wall conditions.

\section{Numerical transition and cosmological conversion}\label{sec:numeric}
The analytic solution identifies the role of finite occupation and supplies useful limits. {In the broken phase, particle masses may become comparable to the temperature.} {We therefore evaluate the complete equilibrium integrals, without relying on the thermal mass truncation.} The numerical treatment follows standard bounce and conversion procedures for scalar theories \cite{Linde1983,AthronReview2024,AthronPhaseTracer2025}, while the additional species contributes only its postulated thermodynamic free energy.

{We introduce} a mass unit $M=100\,\mathrm{GeV}$, a dimensionless field $\varphi=\phi/M$ and temperature $\theta=T/M$. {The subtracted potential is defined as} $
 \calU_q(\varphi,\theta)=
 \frac{V_{\eff}(M\varphi,M\theta)-V_{\eff}(0,M\theta)}{M^4}$.

We specify our numerical bounds by virtue of the following limits:
\begin{subequations}
    \begin{align}
 \calU_0(\varphi)&=-\frac{0.08}{2}\varphi^2
 -\frac{0.22}{4}\varphi^4
 +\frac{0.20}{6}\varphi^6,\label{tree}\\
 m_X/M&=0.55|\varphi|,\\
 m_B/M&=\sqrt{0.80^2+0.24^2\varphi^2},\\
 m_F/M&=\sqrt{0.70^2+0.18^2\varphi^2},\\
 (g_X,g_B,g_F)&=(8,2,4),
 \label{massBenchmark}
\end{align}
\end{subequations}
with subscripts $B$ and $F$ indicating bosonic and fermionic populations. We keep the same phenomenological parameters as the previous analysis. 

The complete potential, adopted in the numerical scan, reads: $
 \calU_q={}\calU_0+\calU_X^{(q)}
 +\calU_B+\calU_F+\calU_{\rm CW}^{R}+\calU_{\rm ring}$, where all pieces vanish at $\varphi=0$ by subtraction. In particular,
\begin{equation}
 \calU_X^{(q)}=
 \frac{g_X\theta^4}{2\pi^2}
 \left[\calJ_q\!\left(\frac{m_X^2}{T^2}\right)
 -\calJ_q(0)\right].\label{UX}
\end{equation}
For conventional degrees of freedom, $\calU_B$ and $\calU_F$ are obtained by subtracting their values at the origin in Eq.~\eqref{fB} and dividing by $M^4$.

{For the conventional species, we evaluate the vacuum contribution using scalar and Dirac determinants, with a finite subtraction at} $\varphi=0$. Write $u=\varphi^2$, $y_i=y_{0i}+k_iu$, $y_{0i}=m_{i0}^2/M^2$ and $k_i=h_i^2$. With $s_B=+1$, $s_F=-1$ and $Q=\mu_R/M$, we have
\begin{subequations}
\begin{align}
 W_i(y)&=\frac{s_i g_i}{64\pi^2}y^2
 \left[\ln\frac{y}{Q^2}-\frac32\right],\\
 \calU_{{\rm CW},i}^{R}(u)
 &=W_i(y_{0i}+k_iu)-W_i(y_{0i})\nonumber\\
 &\quad-k_iuW_i'(y_{0i})
 -\frac{k_i^2u^2}{2}W_i''(y_{0i}).
 \label{CWren}
\end{align}
\end{subequations}
Primes in Eq.~\eqref{CWren} denote derivatives with respect to $y$. We use $Q=1$ and $c_i=3/2$ for these conventional fields. {The subtraction removes the constant, quadratic and quartic contributions to the vacuum potential around the symmetric origin.} At the displayed fixed order, the resulting function has no explicit dependence on $Q$ because the subtracted terms exhaust that dependence. The coefficient choice does not establish renormalization group independence in a full theory, and no unmotivated vacuum determinant is assigned to $X$.

For the conventional boson, we adopt an Arnold-Espinosa type ring term with a phenomenological thermal screening mass $\Pi_B/M^2=\kappa_B\theta^2$,

\begin{align}
 \calU_{\rm ring}
 ={}&-\frac{g_B\theta}{12\pi}\Bigl\{
 \left[(\widetilde m_B^2+\kappa_B\theta^2)^{3/2}
 -\widetilde m_B^3\right]\nonumber\\
 &\quad-\left[(\widetilde m_{B0}^2+\kappa_B\theta^2)^{3/2}
 -\widetilde m_{B0}^3\right]\Bigr\},
 \label{ring}
\end{align}
with $\widetilde m_B=m_B/M$ and
$\kappa_B=0.20$. 

\subsection{Bounce action}

{For each value of} $q${, we determine the symmetric and broken minima, with} $\varphi_c>0${, satisfying} $\partial_\varphi\calU_q(\varphi_c,\theta_c)=0$ and $\calU_q(\varphi_c,\theta_c)=0$.

{We also require a local maximum between the two minima to establish the first order character of the transition.} The degeneracy conditions alone are insufficient if the intermediate extrema have not been checked. {We compute the energy density difference between the phases from}
\begin{subequations}
    \begin{align}
 \Delta\rho(T)&=M^4\left[-\calU_q
 +\theta\,\partial_\theta\calU_q\right]_{\varphi=\varphi_{\true}},
 \label{drhonum}\\ \sigma_c/M^3&=\int_0^{\varphi_c}\!\dd\varphi\,
  \sqrt{2\calU_q(\varphi,\theta_c)}.
 \label{sigmanum}
\end{align}
\end{subequations}

The derivative is taken at fixed $\varphi$ and evaluated along the stationary broken phase. The integral in Eq.~\eqref{sigmanum} uses the degenerate potential and represents its planar interface tension.

For an $O(3)$ symmetric thermal configuration, the Euclidean action is
\begin{equation}
 S_3=4\pi\int_0^\infty\!\dd r\,r^2
 \left[\frac12(\partial_r\phi)^2+\Delta V_{\eff}(\phi,T)\right].
 \label{S3}
\end{equation}
Here $\Delta V_{\eff}=V_{\eff}(\phi,T)-V_{\eff}(0,T)$. Variation gives $
\phi''+\frac2r\phi'=\partial_\phi V_{\eff}$, with $ \phi'(0)=0$ and $\phi(\infty)=0$.

{The boundary value problem fixes the central value of the field.} With $x=Mr$ the dimensionless system reads
\begin{subequations}
    \begin{align}
 \varphi_{xx}+\frac2x\varphi_x
 &=\partial_\varphi\calU_q,\\
 \frac{S_3}{T}&=\frac{4\pi}{\theta}
 \int_0^\infty\!\dd x\,x^2
 \left[\frac12\varphi_x^2+\calU_q\right].
 \label{bouncedimless}
    \end{align}
\end{subequations}
The regular small radius initial expansion is
$\varphi(x)=\varphi_0+\frac{x^2}{6}\partial_\varphi\calU_q(\varphi_0,\theta)+O(x^4)$.
An independent consistency relation follows from rescaling the radius of a stationary solution. In three dimensions the gradient part of the action balances the potential part according to $T_\nuc+3U_\nuc=0$. Consequently,
\begin{equation}
 S_3=\frac{4\pi M}{3}\int_0^\infty
 \!\dd x\,x^2\varphi_x^2.
 \label{virial}
\end{equation}
{We checked the numerical bounce solutions against this identity.} This checks an action and a solution profile internally, but is not a full convergence study across independent solvers.

\subsection{Cosmological expansion and conversion criteria}
A massless Gentile population has $P_X=-f_X^{(q)}(0,T)$, $\rho_X=3P_X$ and
$\rho_X=\frac{g_X\pi^2T^4}{30}(1-K^{-3})$.
For a massive population, the density must be obtained from the full occupation integral,
\begin{equation}
 \rho_X=\frac{g_X}{2\pi^2}\int_0^\infty\!\dd p\,
 p^2E_p n_q(E_p).
 \label{rhoXmassive}
\end{equation}
In the false phase of our benchmark $m_X=0$. We specify an additional spectator bath with $g_{\rm sp}=100$ massless bosonic energy degrees of freedom and compute the conventional massive $B$ and $F$ contributions from their equilibrium distributions. The Friedmann expansion gives
$H^2=\frac{8\pi G}{3}\rho_{\rm ref}$,
with density $
 \rho_{\rm ref}=\frac{\pi^2}{30}g_{\rm sp}T^4
 +\rho_X(0,T)+\rho_B(T)+\rho_F(T)
 +\rho_{\vac}$ and vacuum contribution
$\rho_{\vac}=-M^4\min_\varphi \calU_q(\varphi,0)$.
The latter fixes the true vacuum to have zero energy at zero temperature and assigns the corresponding positive reference vacuum energy to the symmetric phase. The vacuum term is approximately $0.072345M^4$ for this parameter set. We assume $T\propto a^{-1}$ through the short interval in which nucleation and conversion occur. This is a prescribed thermal history which neglects reheating and the backreaction of released latent heat.

We write the thermal nucleation rate as
\begin{equation}
 \Gamma(T)=A(T)e^{-S_3(T)/T},\label{Gamma}
\end{equation}
where the prefactor is an explicit ansatz, chosen as $ A(T)=T^4$.

{We approximate the accumulated number of nucleation events within a changing Hubble volume by}
\begin{equation}
 N_H(T)=\int_T^{T_c}\!\frac{\dd T'}{T'}
 \frac{\Gamma(T')}{H^4(T')}.
 \label{NH}
\end{equation}
We call $T_n$ the solution of $N_H(T_n)=1$. The integral defines an onset criterion; it is not an exact first bubble probability within a fixed causal region. The inverse duration is
\begin{equation}
 \frac{\beta}{H}=T\frac{\dd}{\dd T}
 \left(\frac{S_3}{T}\right)\bigg|_{T_n}.
 \label{beta}
\end{equation}
{We distinguish the strength normalized to the radiation and thermal plasma energy from that normalized to the total reference energy density,
\begin{subequations}
    \begin{align}
 \alpha_{\rm rad}&=\frac{\Delta\rho(T_n)}
 {\rho_{\rm rad+plasma}(T_n)},\\
 \alpha_{\rm ref}&=\frac{\Delta\rho(T_n)}
 {\rho_{\rm ref}(T_n)},\label{alphas}\\
 \rho_{\rm rad+plasma}&=\rho_{\rm ref}-\rho_{\vac}.
    \end{align}
\end{subequations}
The ordinary massive species are included in this chosen thermal plasma denominator. The traditional pure radiation definition can differ mildly, and $\alpha_{\rm ref}$ must not be identified with $\alpha_{\rm rad}$.

A bubble nucleated at $t'$ has comoving radius $r(t,t')=\int_{t'}^t v_w\dd t''/a(t'')$. For independent nucleation centers the false phase survival fraction is $P_\false=e^{-I}$. At constant wall speed $v_w=0.60$, the conversion exponent becomes
\begin{subequations}
\begin{align}
 &I(t)=\frac{4\pi}{3}\int_{t_c}^{t}\!\dd t'\,
\Gamma(t')a^3(t')\left[\int_{t'}^t\!\frac{v_w\dd t''}{a(t'')}\right]^3,\\
 &I(T)=\frac{4\pi}{3}\int_T^{T_c}\!\dd T'\,
 \frac{\Gamma(T')}{H(T')T'^4} \left[v_w\int_T^{T'}\!\frac{\dd T''}{H(T'')}\right]^3.
 \label{Iconversion}
\end{align}
\end{subequations}
The latter identity uses the stated adiabatic cooling law. We adopt $I(T_p)=0.34$ as an operational percolation marker and $I(T_{95})=3$ as a marker for approximately $95\%$ conversion. Since a comoving volume expands, an additional physical volume condition is $
 \frac{\dd}{\dd t}\ln(a^3e^{-I})=3H-\dot I<0$. At this stage, for the sake of completeness, we remark that the distinction between onset, percolation and physical depletion is commonly explored in the literature, see e.g. Refs.  \cite{GuthWeinberg1981,TurnerWeinbergWidrow1992,MegevandRamirez2017,EllisLewickiNo2019}.

\subsection{Numerical results}

{We evaluated the thermal momentum integrals using Gauss-Legendre quadrature with 240 nodes and a dimensionless cutoff of} $42$. Coexistence and stationary points were obtained by root finding and one dimensional minimization. The resulting spinodal temperatures in Table~\ref{tab:coex} satisfy the expected ordering $T_0<T_c<T_{\rm sp}$, and their local derivative and curvature conditions were checked independently by finite differences. The bounce was evaluated with radial overshoot and undershoot shooting, using a cubic spline for the potential and its derivative. For numerical approaches to more general scalar potentials, see Refs.~\cite{KonstandinHuber2006,Wainwright2012}. The $S_3/T$ curve was interpolated with a monotonicity preserving piecewise cubic interpolant before integration of Eqs.~\eqref{NH} and \eqref{Iconversion}. {We considered} $q=1,2,5$ {and the limiting bosonic thermal integral}. {The results in} Tables~\ref{tab:coex} and \ref{tab:nucleation} {refer to this fixed parameter benchmark.}

\begin{table*}[t]
\caption{Coexistence and spinodal data for the complete sextic phenomenological potential, with $M=100\,\mathrm{GeV}$. The lower spinodal $T_0$ is defined by vanishing curvature at the origin and the upper spinodal $T_{\rm sp}$ is the fold of the nonzero extrema at $\phi_{\rm sp}$. The limiting row is the thermal Bose limit with the same fixed vacuum prescription.}\label{tab:coex}
\begin{ruledtabular}
\begin{tabular}{cccccccc}
$q$ & $T_0/M$ & $T_c/M$ & $T_{\rm sp}/M$ & $\phi_c/M$ & $\phi_{\rm sp}/M$ & $\sigma/M^3$ & $L/M^4$\\
1 & 0.85999 & 1.10706 & 1.17806 & 0.93868 & 0.76664 & 0.05080 & 0.11797\\
2 & 0.75278 & 0.98954 & 1.05718 & 0.95436 & 0.77967 & 0.05491 & 0.12495\\
5 & 0.67754 & 0.92423 & 0.99424 & 0.97557 & 0.79735 & 0.06154 & 0.13526\\
$\infty$ & 0.62119 & 0.90866 & 0.98440 & 0.98720 & 0.80521 & 0.06720 & 0.14132
\end{tabular}
\end{ruledtabular}
\end{table*}

\begin{table*}[t]
\caption{Nucleation and conversion outputs for the prescribed expansion history and prefactor $A=T^4$. Temperature entries are in GeV, $\alpha_{\rm rad}$ is normalized to the thermal plasma energy and $\alpha_{\rm ref}$ includes the constant reference vacuum energy. $T_p$ solves $I=0.34$ and $T_{95}$ solves $I=3$. The final column verifies the decrease of the physical false phase volume at $T_{95}$ in the assumed model.}\label{tab:nucleation}
\begin{ruledtabular}
\begin{tabular}{cccccccccc}
$q$ & $T_n$ & $S_3(T_n)/T_n$ & $\beta/H$ & $\alpha_{\rm rad}$ & $\alpha_{\rm ref}$ & $T_p$ & $T_{95}$ & $I(T_n)$ & $\dot I>3H$\\
1 & 106.24 & 137.07 & 5413 & 0.002605 & 0.002601 & 105.74 & 105.69 & $3.0\times10^{-11}$ & yes\\
2 & 94.43 & 137.66 & 4726 & 0.004372 & 0.004362 & 93.94 & 93.88 & $4.5\times10^{-11}$ & yes\\
5 & 87.61 & 138.10 & 4120 & 0.006291 & 0.006271 & 87.09 & 87.04 & $6.9\times10^{-11}$ & yes\\
$\infty$ & 85.66 & 138.28 & 3722 & 0.007097 & 0.007071 & 85.11 & 85.05 & $9.3\times10^{-11}$ & yes
\end{tabular}
\end{ruledtabular}
\end{table*}

\begin{figure}[t]
\centering
\includegraphics[width=\columnwidth]{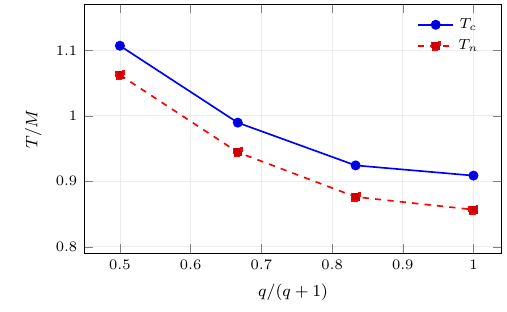}
\caption{Critical and nucleation temperatures against the occupation factor for the fixed numerical benchmark. The right limit denotes the limiting bosonic thermal free energy. Curves connect the discrete calculated points only as a visual guide.}
\label{fig:Tresults}
\end{figure}

\begin{figure}[t]
\centering
\includegraphics[width=\columnwidth]{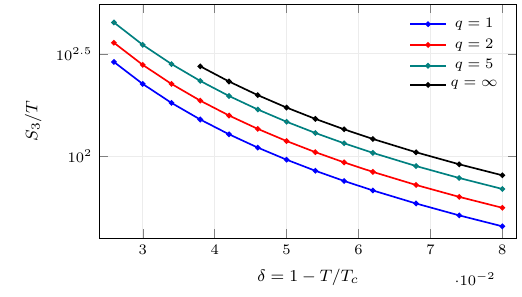}
\caption{Thermal bounce exponent calculated from the complete potential, shown against the fractional supercooling for the same four statistical cases as Tables~\ref{tab:coex} and \ref{tab:nucleation}. Lines connect the actual radial shooting nodes. A logarithmic vertical scale displays the rapid suppression of nucleation close to coexistence.}
\label{fig:bounces}
\end{figure}

\begin{figure}[t]
\centering
\includegraphics[width=\columnwidth]{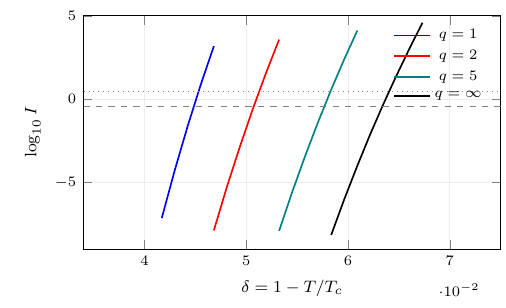}
\caption{Conversion exponent obtained by integrating the calculated thermal bounce rates with the prescribed adiabatic expansion and wall speed. The dashed and dotted horizontal lines mark the operational thresholds $I=0.34$ and $I=3$. The sampling points shown are interpolations of the computed conversion histories, not additional bounce solutions.}
\label{fig:percolation}
\end{figure}

Figure~\ref{fig:Tresults} shows that the critical temperature decreases and the fractional supercooling at onset increases with the allowed occupation number. Figure~\ref{fig:bounces} shows the corresponding bounce exponents, while Fig.~\ref{fig:percolation} displays the conversion histories. The latent heat and coexistence wall tension also increase. {At a common temperature, the free energy profiles remain distinct, showing that the differences cannot be reduced to a uniform rescaling of temperature.} In the numerical sample, both nucleation and the operational conversion conditions are reached. The calculation does not establish an ordering of these quantities for arbitrary couplings or arbitrary scalar potentials.

The virial comparison in Eq.~\eqref{virial} was reported to have a maximum relative discrepancy below $3.6\times10^{-5}$ on the sampled bounce nodes. The critical coexistence conditions were checked numerically. The high temperature mass coefficient and the massless limit were compared against the exact analytic identities. These numerical checks support the internal precision of the published benchmark digits; they do not replace an independent solver comparison, a field dependent kinetic calculation or a complete uncertainty budget.

\section{Physical interpretation and discussion}\label{sec:discussion}

We first discuss statistical limits and origin of the barrier. Hence, the finite occupation law modifies the infrared population of the thermal bath and, through a field dependent mass, its contribution to the scalar free energy. At every fixed finite occupation bound, the cubic mass term cancels in the high temperature domain specified in {Sec.~\ref{sec:background}}. Consequently, the bosonic mechanism associated with the cubic contribution in Eq.~\eqref{boson_exp} is absent from the corresponding Gentile expansion. This result has a direct physical implication for models in which a thermal cubic interaction is responsible for separating competing minima. However, the absence of this particular term must not be identified with the absence of every possible barrier or false vacuum.

The quartic model of Sec.~\ref{sec:q4scan} makes this distinction explicit. Its tree potential has a positive quartic coefficient and does not contain a prescribed barrier. Nevertheless, the finite $q$ thermal function retains a contribution proportional to $\phi^4\ln\phi^2$, as shown in Eq.~\eqref{q4logv}. Within the thermal only prescription, this term can generate a local maximum between the symmetric and broken minima and thus an extremely weak first order transition. The coexistence and spinodal equations provide a direct test of the extrema without assuming their existence. Figure~\ref{fig:q4phase} displays the occupation dependence of the resulting order parameter discontinuity for three values of the quartic coupling. In the reference thermal only benchmark, Table~\ref{tab:q4baseline} gives $\phi_c/T_c\simeq8.52\times10^{-5}$ and a barrier of about $6.22\times10^{-21}T_c^4$ at $q=1$. These exceedingly small values follow from the asymptotic structure of the specified potential, but are not protected against omitted quantum corrections. The figure is a map of equilibrium coexistence strength, not a map of bubble nucleation or cosmological completion. In particular, a nonzero but extremely weak barrier does not by itself establish an observable or dynamically completed first order transition. {The separate bounce and conversion test in Sec.~\ref{sec:q4bounce} locates these operational thresholds for the representative thermal-only models in Table~\ref{tab:q4dyn}, subject to the quasistatic and infrared limitations quantified above.}

The physical interpretation of this residual barrier depends on the vacuum prescription. The quartic scan deliberately includes only the thermal contribution of the $X$ population. At $q=1$, the thermal partition function is exactly fermionic, but the logarithmic field dependence of the isolated thermal free energy is canceled when the conventional fermionic vacuum determinant is included. In the reference quartic benchmark, this completion gives $\lambda_{\mathrm{eff},F}(T_0)\simeq0.09581>0$ [Eq.~\eqref{q4matchedF}], and the scanned transition becomes continuous near $T_0$. Therefore, the weak first order behavior of the thermal only quartic model cannot be promoted to a universal prediction of Gentile statistics. At intermediate occupation bounds, the partition function alone does not determine a unique vacuum determinant or a corresponding renormalization prescription; a microscopic model is required for that extension.

The sextic construction addresses a separate and complementary situation. Its negative quartic and positive sextic scalar interactions supply an independent coexistence barrier. The complete thermal functions then modify the phase minima, coexistence temperatures and wall tension, and thereby the action controlling critical bubbles. The numerical results of {Sec.~\ref{sec:numeric}} show that changing the occupation bound modifies nucleation onset, inverse duration and false vacuum conversion within this prescribed phenomenological model. These results do not require the cubic term of the $X$ bath to generate the barrier.

The formal statistical limits must also be interpreted carefully. The case $q=1$ reproduces the ideal Fermi occupation and thermal free energy, while $q\to\infty$ reproduces the ideal Bose thermal free energy at fixed positive mass and temperature. Neither statement, by itself, determines the complete interacting quantum effective action, its fluctuation determinants or the requisite infrared resummations. The high temperature expansion is nonuniform in $q$: the domain $Km/T\ll1$ shrinks as $K=q+1$ increases. For the broken minima at coexistence in the sextic benchmark, $Km_X(\phi_c)/T_c$ is approximately $0.93$, $1.59$ and $3.48$ for $q=1,2$ and $5$, respectively. We therefore use the complete thermal integrals in the numerical analysis rather than extrapolating the finite $q$ cubic cancellation beyond its domain of validity. The common vacuum prescription used in the sextic comparison fixes a well defined phenomenological benchmark, but it is not equivalent to matching complete fermionic and bosonic field theories at the limits.

\subsection{Sensitivity to the thermal history and screening}

The nucleation rate prefactor, bubble wall velocity and scalar thermal screening are incompletely determined in the sextic benchmark. To assess part of this dependence, we vary $A/T^4$ between $0.1$ and $10$ and $v_w$ between $0.3$ and $1$, obtaining the following ranges for the supercooling parameter $\delta_{95}=1-T_{95}/T_c$:
$0.04445\leq\delta_{95}\leq0.04636$ for $q=1$ and $0.06282\leq\delta_{95}\leq0.06552$ for $q\to\infty$.
These ranges were obtained at fixed bounce curves and with a prescribed adiabatic expansion history. Varying the conventional bosonic ring parameter $\kappa_B$ between zero and $0.5$ changes $T_c/M$ from $1.10132$ to $1.11457$ for $q=1$ and from $0.90618$ to $0.91199$ at the bosonic thermal limit. Within these limited tests, the shifts associated with the occupation bound exceed those produced by this particular screening variation. This comparison is not a complete uncertainty estimate: it does not include a self consistent treatment of the scalar order parameter fluctuations, higher order contributions, derivative corrections or a dynamical calculation of wall friction. Nor do these numerical bounds apply to the separate quartic coexistence scan of Sec.~\ref{sec:q4scan}.

The transition strengths are small and the inverse durations in Hubble units are large for the selected sextic benchmark. The transition is consequently weak and rapid within the assumed background. We do not calculate the hydrodynamic efficiencies, acoustic lifetimes or gravitational wave spectra. These quantities require a treatment of plasma perturbations and wall propagation beyond the equilibrium potential \cite{AthronReview2024,EllisLewickiNo2019,Caprini2020}. {The quartic computation in Sec.~\ref{sec:q4bounce} gives formal nucleation thresholds before the spinodal for the benchmark occupation bounds. The infrared estimates and the prescribed cooling law limit their interpretation as physical nucleation and conversion predictions. Extending the calculation to the full coexistence map remains a separate task.}

\subsection{Cosmological consequences}

If the $X$ population remains ultrarelativistic at neutrino decoupling or recombination, its additional energy density contributes to the effective number of neutrino species. For a thermal bath of temperature $T_X$,
\begin{equation}
 \Delta N_{\eff}=\frac{4g_X}{7}
 \left(1-\frac{1}{(q+1)^3}\right)
 \left(\frac{T_X}{T_\nu}\right)^4.
 \label{Neff}
\end{equation}
Here one reference neutrino species includes the neutrino and antineutrino helicity states. For the chosen $g_X=8$ and $T_X=T_\nu$, the limits give $\Delta N_{\eff}=4$ and $32/7$. These values would represent substantial contributions to the radiation density. Earlier thermal decoupling can dilute $T_X/T_\nu$; for example, the illustrative entropy ratio $[10.75/106.75]^{1/3}$ gives $\Delta N_{\eff}\simeq0.187$ for $q=1$ and $0.214$ at the Bose limit. This estimate assumes the standard visible sector entropy history and a surviving relativistic $X$ population. It is a conditional mapping rather than a fit to data. Observational constraints from cosmic microwave background measurements must be imposed on a specified thermal history \cite{Planck2020}.

In the sextic benchmark, $X$ becomes massive in the broken phase, with a characteristic mass around $55\,\mathrm{GeV}$ at the transition. It need not remain relativistic at Big Bang nucleosynthesis. The resulting abundance, possible annihilation or decay and any entropy transfer depend on interactions not specified by the occupation ansatz. Consequently, neither a relic density prediction nor a BBN exclusion or acceptance follows from Eq.~\eqref{Neff} alone. The quartic phase map likewise does not, by itself, determine the relic abundance or the subsequent thermal history of $X$.

\section{Final outlooks and perspectives}\label{sec:conclusion}

In this work, we investigated thermal bubble nucleation
in a canonical scalar theory coupled to an equilibrium population
obeying Gentile occupation statistics. We derived the free energy
from the finite occupation partition function and obtained an exact
relation to ordinary bosonic thermal integrals. The high temperature
expansion showed that the cubic mass contribution cancels for each
fixed finite occupation bound within its domain of validity.
This removes a familiar bosonic mechanism for generating a thermal
barrier, but does not in itself exclude metastability.
Interestingly, the high temperature expansion is nonuniform
in the occupation bound, enabling us to understand why the cubic
mass contribution is recovered at the bosonic limit.
Further, we found that the thermal mass correction induced by
the Gentile population depends on the occupation bound through
the factor $q/(q+1)$.

To examine the latter possibility, we studied a quartic scalar model
with no prescribed tree level barrier. Solving the stationary point,
coexistence and spinodal conditions with the complete Gentile thermal
integral, we found that a residual quartic logarithm can support a
first order transition when the vacuum contribution of the
bath is omitted. We constructed an occupation dependent
phase diagram and showed that the discontinuity in the
order parameter can be exponentially small, particularly near the
fermionic limit. We also showed that this result depends on the
vacuum prescription: including the conventional fermionic vacuum
contribution changes the phase structure at the fermionic
limit by canceling the logarithmic field dependence
responsible for the residual barrier and removes the weak first
order transition in the neighborhood of the temperature at which
the origin loses its stability.

Accordingly, while the cancellation of the cubic mass
contribution follows directly from Gentile statistics, the existence
of a residual barrier depends on the adopted effective potential.
Within the adopted thermal potential and prescribed thermal
history, we found that the formal nucleation, percolation 
and conversion thresholds occur before the symmetric spinodal
for representative occupation bounds. However, comparing the
characteristic scales of bubble formation and cosmological cooling, as well as the characteristic scalar relaxation time, we found that the quasistatic approximation breaks down at
the fermionic limit.
Further, our estimates of the infrared scalar contributions indicate
a loss of perturbative control for the first three finite occupation
bounds. Consequently, the existence of a static bounce and the
formal fulfillment of the conversion criteria do not necessarily
imply that the transition can be dynamically completed.
We also investigated the effects of the finite initial radius
of the nucleated bubbles, finding negligible corrections to
the conversion history for $q\geq3$ and small corrections for
$q=2$ within our adopted prescription.

We then investigated a sextic potential in which the scalar
self interactions generate a first order barrier independently of
the occupation statistics. We derived its coexistence conditions,
latent heat, wall profile and surface tension analytically.
In particular, within the approximation in which the
Gentile contribution enters through the quadratic thermal mass
correction, we found that both the latent heat and the surface
tension are independent of the occupation bound for fixed
scalar potential parameters.

Using the complete thermal integrals, we numerically
determined the phase minima, thermal bounce action, nucleation
onset and the conditions for false vacuum conversion.
The critical temperatures and nucleation parameters varied
systematically with the occupation bound. We also recovered
the thermal fermionic and bosonic limits within the same
phenomenological prescription.
Unlike the aforementioned analytic approximation,
the complete thermal integrals introduce an occupation
dependence in both the latent heat and the surface tension.

In particular, increasing the maximum occupation number
decreased the critical and nucleation temperatures while increasing
the latent heat, surface tension and fractional supercooling in
our sextic benchmark.
Further, the transition strength increased, whereas
the inverse duration decreased.
Nevertheless, the transitions remained weak and
rapid throughout the numerical cases investigated.
The differences among the corresponding free energy profiles
at a common temperature indicate that the effects of Gentile
statistics cannot be accounted for by simply rescaling
the temperature.

Thus, we found
that the nucleation, percolation,  conversion
criteria, etc. are fulfilled in all the sextic cases investigated.
In addition, the physical volume occupied by the false vacuum
decreases at the adopted conversion threshold. We also examined
the sensitivity of these results to variations in the nucleation
rate prefactor, bubble wall velocity and thermal screening
of the conventional bosonic sector. In particular, the shifts
associated with the occupation bound exceed those obtained
by varying the conventional bosonic screening parameter
over the considered interval. 

The quartic and sextic analyses therefore address complementary
questions: whether metastability can survive the removal of the
cubic thermal term and how finite occupation statistics
modify a cosmological transition whose barrier has an independent
origin.

From a theoretical perspective, a further development
of our approach would require a microscopic description of the
statistical sector. While the Gentile occupation law defines a
thermal free energy, it does not uniquely determine vacuum
fluctuations, microscopic interaction vertices or thermal self
energies. The different phase structures obtained at the
fermionic limit with and without the vacuum contribution
illustrate this limitation.
A consistent comparison of the limits should include the
renormalized vacuum action and the appropriate infrared
resummations in the bosonic case, with the same perturbative
accuracy in the scalar sector. For intermediate occupation bounds,
a dynamical realization of the Gentile sector would be
needed to determine the corresponding vacuum terms, equilibration
rates and possible gradient corrections.

In this respect, increasing the occupation bound improves
the local perturbative expansion of the scalar infrared sector,
although perturbative control along the entire bounce profile
remains to be investigated.
For sufficiently small occupation bounds, a
nonperturbative treatment of the scalar sector appears
particularly relevant, given the enhancement of infrared
corrections associated with the exponentially small
discontinuity in the order parameter.
A wider scan is needed to distinguish
regions of successful nucleation, spinodal conversion and incomplete
transitions for other occupation bounds and couplings. For a
dynamical study of spinodal conversion in a different microscopic
framework, see Ref.~\cite{Bea2025}.
Such an investigation should also account for uncertainties
associated with the rate prefactor, scalar thermal screening,
vacuum prescription, numerical convergence and higher order
operators in the effective potential. Quantum tunneling may also
become relevant where thermal nucleation is suppressed at lower
temperature.

Finally, to extend our results to a self consistent
cosmological scenario, one should account for the dynamics of
the plasma and determine the corresponding cooling law and
bubble wall velocity. Entropy production, reheating, fluid
backreaction and wall friction could change the duration and
completion of the transition.

In particular,  reheating could become relevant
for the weak quartic transitions investigated here, owing to
the narrow temperature interval over which false vacuum
conversion is expected to occur.

With the corresponding hydrodynamic
efficiencies and source lifetimes specified, the stochastic
gravitational wave background could be calculated.
The weak and rapid transitions obtained in our sextic
benchmark do not suggest a strong gravitational wave signal,
although different regions of the parameter space may lead to
different outcomes.

Additionally, we derived
the contribution of a relativistic Gentile population to
the effective number of neutrino species, finding an explicit
dependence on both the occupation bound and the temperature
of the additional sector. Assuming an early thermal decoupling
and a surviving relativistic population, we showed that
the corresponding contribution can be substantially diluted
by the subsequent entropy evolution of the conventional
plasma. However, since the particles belonging to the
Gentile sector acquire a mass in the broken phase of our
benchmark, their contribution to the radiation density
at later cosmological epochs cannot be inferred from the
occupation law alone.

Additional constraints would depend on the Gentile particle microphysics,
including their contribution to the radiation density during
primordial nucleosynthesis and recombination.
We leave these investigations to future work, with the
purpose of determining whether deviations from the conventional
occupation statistics can lead to observable effects in the
thermal history of the universe.

\begin{acknowledgments}
OL thanks Roberto Della Ceca and the Brera National Institute of Astrophysics for support and acknowledges the Istituto Nazionale di
Fisica Nucleare (INFN) through the Iniziativa Specifica
QGSKY. OL is also grateful to the students of Particle Cosmology course at the University of Camerino, with whom he had the pleasure to discuss about nucleation in the early universe with alternative statistics. The work of HQ was partially supported by PAPIIT-DGAPA-UNAM, grant No.~108225, and by CONAHCYT, grant No.~CBF-2025-I-253. HQ is also grateful to the University of Camerino for his stay during which this paper has been written and to the INFN, sezione di Perugia, for economical support.

\end{acknowledgments}

\appendix
\section{Statistical limits and dimensional checks}
\label{app:checks}

For $K=2${,} the geometric sum satisfies $(1-x^2)/(1-x)=1+x$; thus
$\calJ_{q=1}(y)=-J_F^+(y)$, with $J_F^+(y)=\int_0^\infty\!\dd x\,x^2\ln(1+e^{-\sqrt{x^2+y}})$.
For fixed $y>0$, $K^{-3}J_B(K^2y)$ vanishes as $K\to\infty$. At fixed finite $K$, the massless free energy {agrees with} Eq.~\eqref{HTgentile}. {The free energy densities} $f_X$ and $V_{\eff}$ {have mass dimension four, while} $S_3$ {has mass dimension one,} $\Gamma$ {has mass dimension four and} $\sigma$ {has mass dimension three}. The combinations $S_3/T$, $I$, $N_H$, $\alpha$ and $\beta/H$ are dimensionless.

\section{Bounce and cosmological conversion checks}

At $T\lesssim T_c${,} the thin wall approximation gives
$\frac{S_3}{T}\simeq\frac{16\pi\sigma_c^3}{3TL^2(1-T/T_c)^2}$.
{The exact bounce is required when the wall thickness becomes comparable to the bubble radius.} For a stationary solution{,} {we introduce} $S_3(\lambda)=\lambda^{-1}T_g+\lambda^{-3}U_g$ {through} the transformation $\phi_\lambda(r)=\phi(\lambda r)$. Since $\dd S_3/\dd\lambda$ vanishes at unity, $T_g+3U_g=0$, where $T_g=2\pi\int r^2\phi'^2\dd r$. {We thus recover} Eq.~\eqref{virial}, {which provides an internal check of the numerical action}.

Now, introducing $\delta=1-T/T_c$ and $\delta'=1-T'/T_c$, {we rewrite} Eq.~\eqref{Iconversion} {as}
\begin{equation}
\begin{aligned}
 I(\delta)=\frac{4\pi}{3}\int_0^\delta\!\dd\delta'\,
 \frac{T_c\Gamma[\delta']}{H[\delta']T(\delta')^4}\times\left[v_w\int_{\delta'}^\delta\!\frac{T_c\dd\xi}{H[\xi]}\right]^3,
\end{aligned}
\label{Idelta}
\end{equation}
with $T(\delta)=T_c(1-\delta)$. 
{The decrease of the physical false vacuum volume follows from} $a^3P_\false=a^3e^{-I}$ {and is independent of the operational percolation threshold}.


\begin{thebibliography}{99}
\bibitem{Langer1969} J.~S.~Langer, Ann. Phys. (N.Y.) {54}, 258 (1969).
\bibitem{Affleck1981} I.~Affleck, Phys. Rev. Lett. {46}, 388 (1981).
\bibitem{Coleman1977} S.~Coleman, Phys. Rev. D {15}, 2929 (1977).
\bibitem{CallanColeman1977} C.~G.~Callan and S.~Coleman, Phys. Rev. D {16}, 1762 (1977).
\bibitem{Linde1983} A.~D.~Linde, Nucl. Phys. B {216}, 421 (1983).
\bibitem{GuthWeinberg1981} A.~H.~Guth and E.~J.~Weinberg, Phys. Rev. D {23}, 876 (1981).
\bibitem{AthronReview2024} P.~Athron, C.~Bal{\'a}zs, A.~Fowlie, L.~Morris, and L.~Wu, Prog. Part. Nucl. Phys. {135}, 104094 (2024).
\bibitem{ColemanWeinberg1973} S.~Coleman and E.~Weinberg, Phys. Rev. D {7}, 1888 (1973).
\bibitem{DolanJackiw1974} L.~Dolan and R.~Jackiw, Phys. Rev. D {9}, 3320 (1974).
\bibitem{Dine1992} M.~Dine, R.~G.~Leigh, P.~Huet, A.~D.~Linde and D.~A.~Linde, Phys. Rev. D {46}, 550 (1992).
\bibitem{ArnoldEspinosa1993} P.~Arnold and O.~Espinosa, Phys. Rev. D {47}, 3546 (1993).
\bibitem{Gentile1940} G.~Gentile Jr., Nuovo Cimento {17}, 493 (1940).
\bibitem{DaiXie2004} W.-S.~Dai and M.~Xie, Ann. Phys. (N.Y.) {309}, 295 (2004).
\bibitem{DaiXie2013} W.-S.~Dai and M.~Xie, Ann. Phys. (N.Y.) {332}, 166 (2013).
\bibitem{Haldane1991} F.~D.~M.~Haldane, Phys. Rev. Lett. {67}, 937 (1991).
\bibitem{Wu1994} Y.-S.~Wu, Phys. Rev. Lett. {73}, 922 (1994).
\bibitem{Megevand2004} A.~M\'egevand, Phys. Rev. D {69}, 103521 (2004).
\bibitem{Kajantie1996} K.~Kajantie, M.~Laine, K.~Rummukainen and M.~Shaposhnikov, Nucl. Phys. B {458}, 90 (1996).
\bibitem{Gould2021} O.~Gould, JHEP {04}, 057 (2021).
\bibitem{GouldHirvonen2021} O.~Gould and J.~Hirvonen, Phys. Rev. D {104}, 096015 (2021).
\bibitem{MooreRummukainen2001} G.~D.~Moore and K.~Rummukainen, Phys. Rev. D {63}, 045002 (2001).
\bibitem{MooreTranberg2001} G.~D.~Moore, K.~Rummukainen and A.~Tranberg, JHEP {04}, 017 (2001).
\bibitem{GrojeanServantWells2005} C.~Grojean, G.~Servant, and J.~D.~Wells, Phys. Rev. D {71}, 036001 (2005).
\bibitem{AthronPhaseTracer2025} P.~Athron \emph{et al.}, Eur. Phys. J. C {85}, 559 (2025).
\bibitem{TurnerWeinbergWidrow1992} M.~S.~Turner, E.~J.~Weinberg and L.~M.~Widrow, Phys. Rev. D {46}, 2384 (1992).
\bibitem{MegevandRamirez2017} A.~M\'egevand and S.~Ram\'irez, Nucl. Phys. B {919}, 74 (2017).
\bibitem{EllisLewickiNo2019} J.~Ellis, M.~Lewicki, and J.~M.~No, JCAP {04}, 003 (2019).
\bibitem{KonstandinHuber2006} T.~Konstandin and S.~J.~Huber, JCAP {06}, 021 (2006).
\bibitem{Wainwright2012} C.~L.~Wainwright, Comput. Phys. Commun. {183}, 2006 (2012).
\bibitem{Caprini2020} C.~Caprini \emph{et al.}, JCAP {03}, 024 (2020).
\bibitem{Planck2020} Planck Collaboration, Astron. Astrophys. {641}, A6 (2020).
\bibitem{Bea2025} Y.~Bea \emph{et al.}, JHEP {11}, 093 (2025).
\end{thebibliography}
\end{document}